# Untangling cross-frequency coupling in neuroscience

Juhan Aru[1,2], Jaan Aru[1,3,4,5,10], Viola Priesemann[1,3,6], Michael Wibral[7], Luiz Lana[1,3,4,8], Gordon Pipa[9], Wolf Singer[1,3,4], Raul Vicente[1,3,4,7,10*]

[1]Max-Planck Institute for Brain Research, Frankfurt am Main, 60528 Germany

[2]École Normale Supérieure de Lyon (UMPA), Lyon, 69364 France

[3]Frankfurt Institute for Advanced Studies, Frankfurt am Main, 60438 Germany

[4]Ernst Strüngmann Institute, Frankfurt am Main, 60528 Germany

[5]Institute of Public Law, Tartu University, Tallinn, 10119 Estonia

[6]Max-Planck Institute for Dynamics and Self-Organization, Göttingen, 37077 Germany

[7]MEG Unit, Brain Imaging Center, Goethe University, Frankfurt am Main, 60528 Germany

[8]Brain Institute, Federal University of Rio Grande do Norte, Natal, 59056, Brazil

[9]Institute of Cognitive Science, University of Osnabrueck, 49069, Germany

[10]Institute of Computer Science, University of Tartu, Tartu, 50409, Estonia

*Corresponding author: raulvicente@gmail.com

## Abstract

Cross-frequency coupling (CFC) has been proposed to coordinate neural dynamics across spatial and temporal scales. Despite its potential relevance for understanding healthy and pathological brain function, the standard CFC analysis and physiological interpretation come with fundamental problems. For example, apparent CFC can appear because of spectral correlations due to common non-stationarities that may arise in the total absence of interactions between neural frequency components. To provide a road map towards an improved mechanistic understanding of CFC, we organize the available and potential novel statistical/modeling approaches according to their biophysical interpretability. While we do not provide solutions for all the problems described, we provide a list of practical recommendations to avoid common errors and to enhance the interpretability of CFC analysis.

**Highlights**

Fundamental caveats and confounds in the methodology of assessing CFC are discussed.

Significant CFC can be observed without any underlying physiological coupling.

Non-stationarity of a time-series leads to spectral correlations interpreted as CFC.

We offer practical recommendations, which can relieve some of the current confounds.

Further theoretical and experimental work is needed to ground the CFC analysis.

# Cross-frequency coupling: How much is that in real money?

One of the central questions in neuroscience is how neural activity is coordinated across different spatial and temporal scales. An elegant solution to this problem could be that the activity of local neural populations is modulated according to the global neuronal dynamics. As larger populations oscillate and synchronize at lower frequencies and smaller ensembles are active at higher frequencies [1], cross-frequency coupling would facilitate flexible coordination of neural activity simultaneously in time and space. In line with this proposal, many studies have reported such cross-frequency relationships [2-4]. Especially phase-amplitude CFC, where the phase of the low frequency component modulates the amplitude of the high frequency activity, has been claimed to play important functional roles in neural information processing and cognition, e.g. in learning and memory [4-8] Furthermore, changes in CFC patterns have been linked to certain neurological and mental disorders such as Parkinson's disease [9-11], schizophrenia [12-14] and for example social anxiety disorder [15]. Therefore, CFC is potentially essential for normal brain function and understanding of CFC patterns can be crucial for diagnosing and eventually treating various disorders.

The classical analysis of CFC seems very straightforward (Figure 1) and is widely used. However, not all signatures of CFC as detected by this analysis method need to be due to interactions between different physiological processes occurring at different frequencies, as is commonly reported. It has been previously shown that signals with abrupt changes lead to spurious CFC results [16] (see *Supplementary results: Examples of spurious CFC* for a related example). The roots of this problem are much more general. Let us take as an example the Van der Pol oscillator, which is a very simple non-linear relaxation oscillator. Conducting the CFC analysis on this oscillator would indicate that the phase of the low frequency components modulates the activity of the higher frequencies. However, despite strong CFC signal there is no simple physical interpretation for the different frequency components of the oscillator, and even less for their interaction. Indeed, any interpretation in terms of modulating or causally interacting frequencies is misleading as the spectral correlations are related to the non-linear characteristics of a single oscillator (see *Supplementary results: Examples of spurious CFC* for a thorough description of this example).

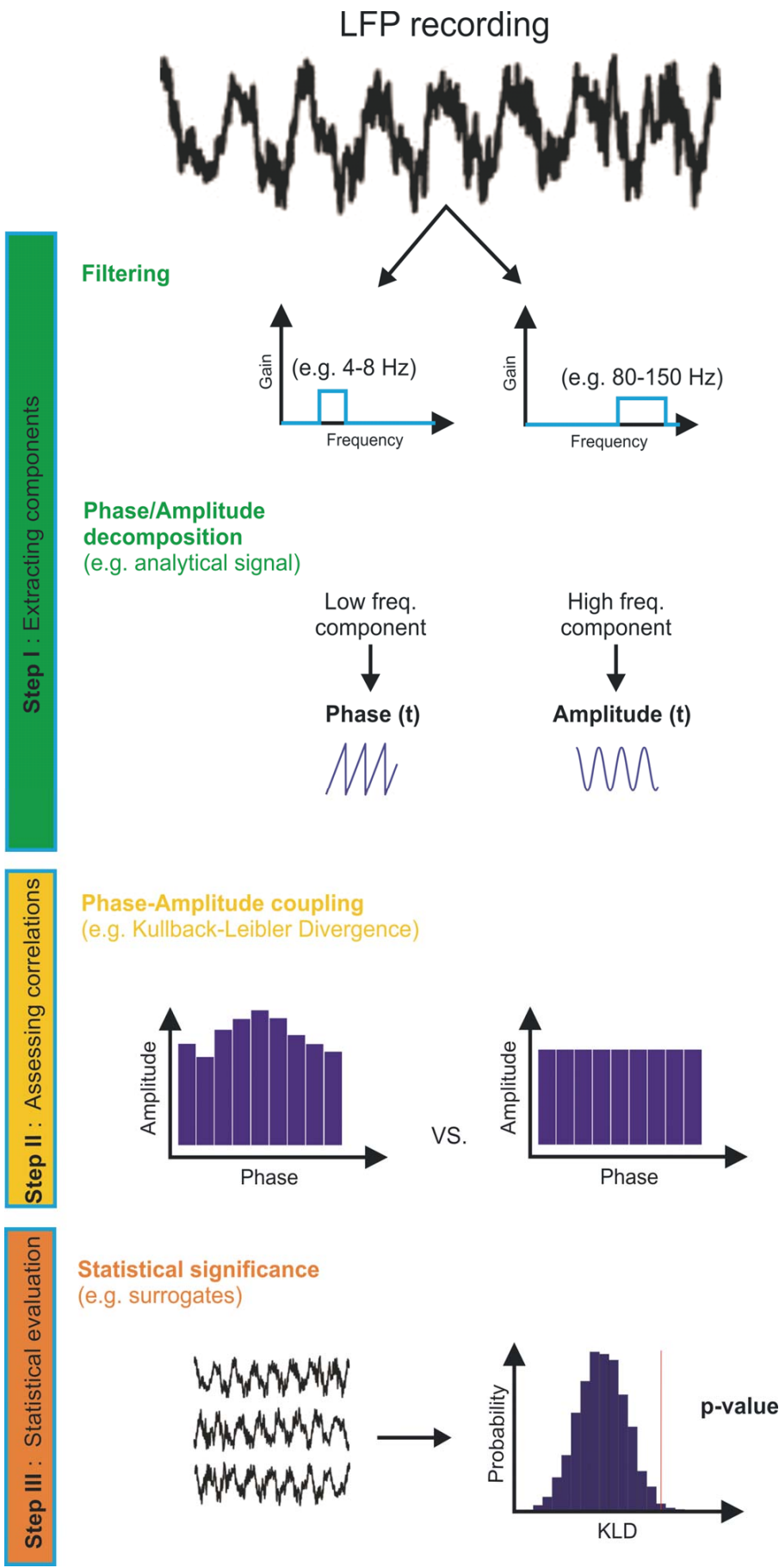

**Figure 1.**
Typical approach to analyze phase-amplitude cross-frequency coupling.

Step 1
Extracting the relevant components. This step is implemented by band-pass filtering and extraction of phase and amplitude dynamics for the relevant frequency bands.

Step 2
Assessing correlations between components. This stage requires the computations of appropriate correlation or dependency measures between amplitude and phase. Therefore, a general measure of phase-amplitude coupling is to precisely quantify how much the histogram of mean amplitude versus phase deviates from a uniform distribution.

Step 3
Statistical evaluation. Parametric or non-parametric approaches comparing to suitable surrogate data can be used to assign a p-value to the observed coupling strength.

This hints that the current analysis of CFC is inherently ambiguous regarding the nature and origin of the observed correlations between the frequency components. A significant CFC measure can be observed in case there are true modulations between subsystems oscillating at different frequencies.

However, it can be also observed under very generic conditions that imply no coupling. Similarly to the example above, any non-linear response where fast components are short-lived compared to the slow components of the signal would produce a significant CFC. In particular this means that current CFC measures of phase-to-amplitude coupling are not specific enough for one to automatically conclude, as it is almost invariably done in the literature, that the phase of a low frequency oscillation modulates the power of high-frequency activity. The same holds for CFC measures of amplitude-to-amplitude, or phase-to-phase coupling.

We wish to emphasize that we do not question the possible importance of CFC as a phenomenon. In fact, we believe that such a mechanism would be an elegant solution to several computational demands the brain has to cope with [1]. However, precisely because of the potential relevance of CFC for understanding the healthy and the pathological brain it is necessary to be aware of the pitfalls and misinterpretations in the methodology currently applied. We hope that the careful assessment of concerns will eventually strengthen the power of CFC analysis as an experimental tool.

The outline of the rest of the paper is as follows: First, we shall point out fundamental caveats and confounds in the current methodology of assessing CFC. Some of these points are original and some others have been known in other fields for years, yet all share the characteristic of being unattended in many of the current studies. Our literature review of the phase-amplitude CFC studies from the years 2010-2014 shows that these issues are relevant and timely (see *literature review* in the *Supplementary material*). Second, we propose an organization of different approaches to CFC analysis according to their biophysical interpretability and statistical inference approach. Finally, we outline some practical recommendations for CFC analysis.

In this Opinion article we can not offer solutions for all the problems of CFC analysis and interpretation that are described, but our hope is that alerting the community to these problems will eventually lead to novel solutions.

# Caveats and confounds of the CFC analysis

In this section we concentrate on what we call the classical CFC analysis – it is illustrated and explained in Figure 1. Any result of this analysis can be used to classify different conditions but only as a marker that is devoid of concrete and clear physiological interpretation. To give a physiological interpretation to CFC, one needs to know the set of potential mechanisms responsible for neural coupling. This set of mechanisms is only beginning to emerge (discussed below). We now discuss some main methodological confounds that make it difficult to build connections between the CFC measure and the underlying neurophysiological processes. More caveats and confounds and examples of spurious CFC can be found in the *Supplementary results*.

### *Instantaneous phase and amplitude: when are they meaningful and when not?*

Standard phase-amplitude CFC analysis proceeds by first selecting two frequency bands followed by the computation of some index for the correlation or dependency between the phase of one band and the amplitude of the other (Fig. 1). The phase and amplitude values extracted from filtered signals can unfortunately only be interpreted in a meaningful way, i.e. as representing physiological oscillations, if a number of basic requirements are met. The same holds for CFC analyses based on them. In *Supplementary discussion: Conditions for a meaningful phase* we present a short but rather thorough review of the conditions that must be met for a meaningful interpretation of phase and amplitude values. The main conclusion is – not that surprisingly - that a clear peak in the power spectrum of the low frequency component is a prerequisite for a meaningful interpretation of any CFC pattern. Our literature review shows that even these well-known conditions were and are not always met in the literature, resulting in a strong over-interpretation of phase and amplitude (see *literature review* in the *Supplementary material*).

### *The importance of the bandwidth*

The two components entering a phase-amplitude CFC analysis after filtering the signal, are determined by the center frequencies and bandwidths of the filters used to isolate them. Our literature review shows that majority of studies proceed by scanning the center frequencies for the phase and amplitude components while keeping a fixed bandwidth of a few Hz. However, this choice of bandwidth is important because it defines what is considered as a component and how the component´s power or group phase changes in time (Figure 2). Thus, it is not the same thing to scan a center frequency from 20 Hz to 60 Hz with a bandwidth of 2 Hz or to consider at once the band centered at 40 Hz with a width of 21 Hz – different effects will be observed. Unfortunately, little or no justification is given to the choice of parameters in most analyses. The choice of bandwidth for the phase component is constrained by the condition of having a meaningful phase and is therefore often correctly chosen to be narrow. However, one also needs to be careful with the bandwidth size for amplitude - if the bandwidth of the higher frequency component ($f_2$) does not include the side peaks produced by the lower frequency ($f_1$), then CFC cannot be detected even if it is present (Figure 2A). Thus certain parameter values usually chosen in the literature can bias the CFC measures towards obtaining false negative results (see *Supplementary Discussion: The importance of the bandwidth*). See also [17].

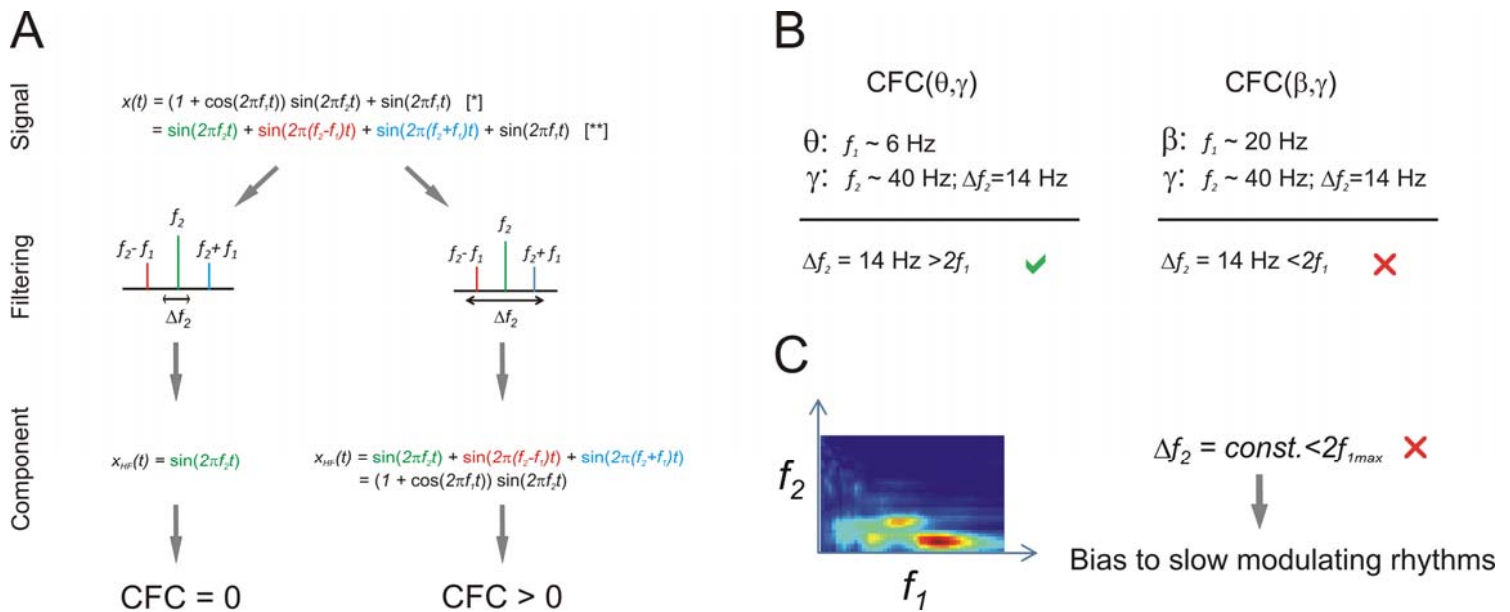

**Figure 2.**
Mathematical decomposition and filtering bandwidth are key parameters to infer and interpret the presence of CFC. A) (Signal) The very same signal can be decomposed into different mathematically equivalent representations ([*] or [**]). The choice of the representation leads to different interpretations regarding the interactions of the components (Filtering and Component). Different filtering bandwidths around the same frequency can lead to different results depending on whether the bandwidth includes modulating sidebands or not. B) For a fixed bandwidth of the modulated frequency, only a range of modulating frequencies can be captured. For example, in the simplified harmonic case, a bandwidth of 14 Hz around a frequency of 40 Hz would allow detection of a potential modulation from a 6 Hz rhythm, but not from a 20 Hz oscillation. C) When scanning the modulating ($f_1$) and modulated ($f_2$) frequencies, a fixed bandwidth biases CFC analysis and favors low frequencies for $f_1$.

### *Non-stationarity and spectral correlations: two sides of the same coin*

Most neuronal signals that we measure are non-stationary. Time-varying sensory stimuli, top-down influences, neuromodulation, endogenous regulatory processes and changes in global physiological states render neuronal dynamics non-stationary. In contrast to a stationary process, a non-stationary process in general exhibits spectral correlations between components of its Fourier expansion [18]. These correlations may be misinterpreted as CFC. The underlying reason for these spectral correlations is that in constructing the spectrum, we decompose a process which is by definition not time-invariant (non-stationary) into the eigenvectors of the time-shift operator, i.e., the complex exponentials in the Fourier expansion. Therefore, in the non-stationary case, there are two possible scenarios leading to positive CFC measures:

One scenario is that physiological processes indeed interact. This interaction then leads to non-stationarities, and at the same time we observe spectral correlations in the Fourier representation. For example, if the phase of a neural input oscillating at theta frequency modulates the amplitude of local gamma oscillations, both obtained from the same LFP recording, the statistical properties of the gamma oscillation amplitude series will change in time, as does theta phase. Specifically, their properties will vary in time only to be repeated after a full cycle of the slow oscillation, and thus exhibiting a particular type of non-stationarity called cyclo-stationarity [19].

The other and problematic scenario is that *unspecific* non-stationarities (that is, *any kind of* change of the statistical properties of the signal), not related to or caused by coupling of neural processes, will also be reflected in spectral correlations which could be over-interpreted as the result of causal interactions among frequency specific neuronal processes. This second scenario can occur if non-stationary input to a given area simultaneously affects the phase of a low frequency component and increases high-frequency activity (common drive to different frequency components of the same signal). For example, typical evoked potentials affect a broad range of frequency components [20]. In this case, high-frequency amplitude increases occur preferentially for certain phases of slow oscillations even without any need of interaction between the two rhythms.

Hence, non-stationary input to a given area can generate correlations between bands, which are not necessarily a signature of interactions between these bands. The argument goes well beyond the relationship between sensory stimulation and CFC in sensory areas: if a brain area under a recording electrode receives time-varying input from any other brain area, this input might generate similar dependencies across frequency components (Figure 3A). The problem is that usually one has no control over the timing of the internal input to the examined brain area (Figure 3B). If this internal input leads to an increase of phase locking for lower frequencies (Figure 3C, left) and at the same time elicits an increase in power at higher frequencies (Figure 3C, middle) phase-amplitude coupling will be observed (Figure 3C, right). The combination of increased activity at high-frequencies and phase-locking to the stimulus of lower frequencies is sufficient to obtain significant measures in standard CFC analysis. Thus, phase-amplitude coupling measured anywhere in the brain can be potentially explained by common influence on the phase and amplitude, without the phase of a low frequency oscillation modulating the power of high frequency activity. In the supplementary materials we illustrate this scenario with examples from the electroretinogram and LFP recorded in the optic tectum of a turtle (Pseudemys scripta elegans) and human intracranial recordings (Supplementary Figures 3 and 4).

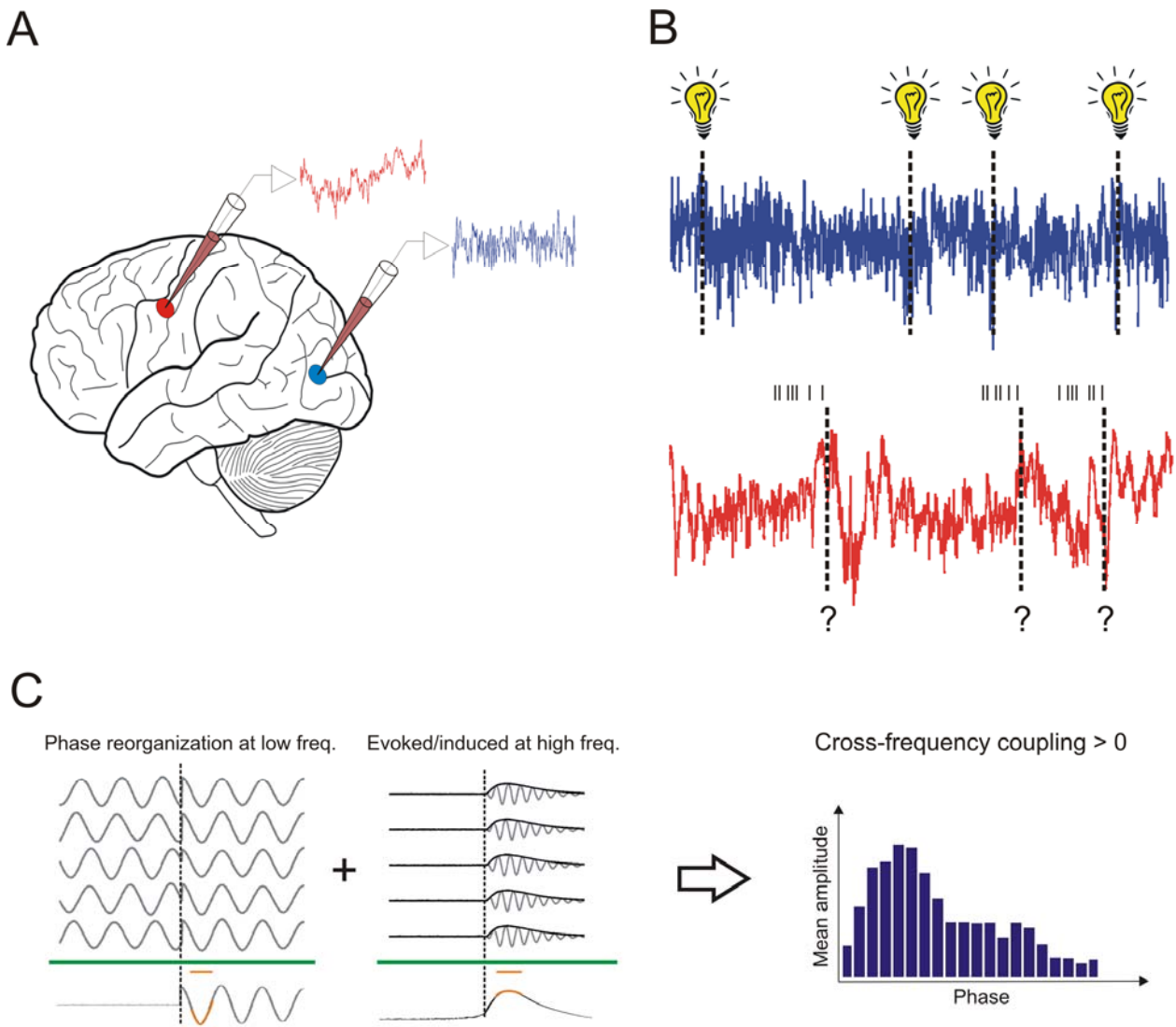

**Figure 3.**
Illustration of how time-varying input can lead to false-positive CFC. A) Illustration of recordings in a cortical sensory area (blue) and a higher cortical area (red). B) Importantly, both can be subject to non-stationary neuronal input but its timing can only be determined for the sensory area. C) Knowing the input timing is necessary to perform phase locking analysis, distinguish between evoked/induced responses and to disambiguate the origin of CFC. Without this additional information the results of the CFC analysis remain ambiguous.

Therefore, the key issue is to distinguish whether the observed phase-amplitude correlation between two bands is due to common drive, generated by external or internal input or whether the correlation is due to a causal interaction between rhythms (which, of course, could also be triggered by the input). Recently, a new approach [21] has been developed to measure transient phase-amplitude coupling directly in an event-related manner. Whereas ideally, their approach of analyzing phase-amplitude relations with respect to the stimulus onset should avoid some event-related artifacts, it is questionable whether the marker actually works as intended (see *Supplementary results:Phase-amplitude coupling for event-related potentials*). Ultimately solving these questions requires a formal causal analysis between the spectral variables of different bands (see also *Supplementary discussion: Causality methods*).

Analysis of between-channel phase-amplitude coupling [22,23] is less likely to be the result of a driving input to a single area. In this research intracranial human data was used to identify the spatial maps of the low frequency (phase providing) and high frequency (amplitude providing) components. From the size and other characteristics of these maps one can conclude that the low- and high- frequency components are separable in the brain. This result is important as such between-channel CFC cannot be created by non-stationary input to one area. However, these findings do not fully solve the underlying problem as these different generators could still be "coupled" by a common driver influencing both generators, rather than by a direct interaction between them.

In conclusion, the above considerations imply that the current phase-amplitude CFC measure is constitutive for the non-stationary responses of driven systems and therefore is not a very specific marker of biophysical coupling. From a mathematical perspective the key aspect is that any consistent response to input, whatever its shape, implies a certain phase locking between its different Fourier components [24,25]. Thus, if the power of any of the fast components lasts a bit more or a bit less than the period of a slow component, then its amplitude will accumulate preferably at certain phases of the slow component. This is all that is needed to give rise to phase-amplitude CFC, as measured for example by the modulation index [26]. It is therefore necessary to recognize that beyond the phase-amplitude CFC index, which is just an index at the signal level, additional information about how the process under study reacts to input and the statistics of the input itself are needed to better resolve the origin of such correlations. Analyses of surrogate data can help to remove some of the ambiguity but only offers partial remedies to the problem, as we shall discuss next.

***Surrogate data: none are perfect but some are better than others***

After some index of CFC has been estimated one needs to rely on statistical inference to reach a conclusion about the statistical significance of the measure. Currently, most studies of CFC rely on the frequentist approach of using surrogate data to estimate a p-value. Some issues related to generation of the surrogate data are discussed here. For the state of some Bayesian approaches see Figure 4 in the section *"Organization of modeling/statistical approaches to CFC"*.

A suitable surrogate construction should only destroy the specific cyclo-stationarities related to the hypothesized CFC effect, while keeping all the unspecific non-stationarities and non-linearities of the original data. Often, it is impossible to construct perfect surrogates that selectively destroy the effect of interest but some approaches are more conservative than others. In our context a sensible requirement is to construct surrogates that minimize the distortion of both phase and amplitude dynamics for each frequency component. If data are organized in detectable repetitive events such as trials locked to an external stimulus or to saccades, shuffling the full/intact phase or amplitude components between different events seems the most straightforward approach. Unfortunately, the very existence of an event-related potential implies that some frequency components are both locked to the event and between themselves. Consequently, this strategy to obtain surrogates alone cannot discern the source of modulation. Future developments of methods to partial out the common drive effect of the event could help to test the significance of a direct phase-amplitude modulation.

Finding appropriate surrogate data for a single continuous stream of data comes with its own challenges. For example, phase scrambling does not meet the minimal distortion criterion, as the generated surrogate data are fully stationary after scrambling, i.e. the non-stationarities of interest and the unspecific non-stationarities are both destroyed alike [27]. A significantly larger CFC in the original data may in this case be due to the removal of non-stationarities not specifically related to physiological CFC. Another approach has been using block re-sampling [3] where one of the continuous time series (i.e., the instantaneous phase) is simultaneously cut at several points and the resulting blocks permuted randomly. This method suffers again from destroying in excess the non-stationary structure of the original series. More conservative surrogates can be obtained by minimizing the number of blocks by cutting at single point at a random location and exchanging the two resulting time courses [3]. Repeating this procedure leads to a set of surrogates with a minimal distortion of the original phase dynamics.

Thus, while perfect surrogate data that selectively disrupt phase-amplitude coupling might be impossible to build (as is the case for most types of non-linear interactions) conservative approaches that minimize distortion of phase and amplitude dynamics can reduce the number of false positives.

### *CFC modulation across conditions*

Several studies have reported significant changes in phase-amplitude CFC with variations of experimental parameters or across two different conditions. The modulation of CFC by the task or experimental condition has then been taken as an indication of its physiological role [4,5,7,28]. However, for now there is only little reason to believe that this modulation could not be due to side-effects of more basic changes between the conditions.

Since the power of bands directly influences the range within which they can modulate or be modulated, it is possible that changes in CFC correlations are a direct consequence of changes in power spectra. For example, changes in the observed CFC can have their origins in the fact that power changes affect the signal-to-noise ratio of phase and amplitude variables and their correlations (e.g.[29]). It is thus necessary to control whether correlations between CFC and other behavioral or physiological variables might be simply due to changes in, for example, the strength or frequency of oscillations. Unfortunately, our literature review shows that in around half of the reviewed studies where conditions are compared, changes in the power spectrum across the conditions are not considered. If the data permit, it is therefore highly recommended to rely on stratification techniques (e.g. [30]) to obtain a subset of matched trials in which the distribution of power across trials is identical for both the phase and amplitude frequency bands in the two conditions to be compared.

In general, as CFC is a statistic based on the correlation of certain variables, it is necessary to control for the explanatory power of these variables themselves, before a specific role for the correlation might be distilled.

## Organization of modeling/statistical approaches to CFC

Until now we have focused on what we call the classical approach (see Figure 1) to assess phase-amplitude CFC effects consisting of: i) isolating frequency components, ii) assessing their dependencies, and iii) computing p-values based on surrogates. We have described how difficult it is at this stage to draw any conclusions about the biophysical mechanisms underlying these measures. However, different frameworks exist to assess relationships among rhythmic processes from experimental time series. A short description of some frameworks can be found in the *Supplementary discussion*. For the purpose of understanding the role of different frameworks in gaining physiological understanding of CFC, we have found it useful to organize them according to their biophysical interpretability and statistical inference approach (Figure 4).

We believe the location of the method along those axes (Figure 4) has to be taken into account to avoid over-interpretations of the results of a CFC analysis. The first section of this paper can be in fact seen as an explanation as to why some models including the classical approaches are positioned very close to the "marker" section. For example, a simple correlation-based quantifier as provided by classical approaches is probably all that is needed, if the sole purpose of CFC analysis is to have a marker to classify different conditions (e.g. disease states). If one insists that this marker should be more specific than say just changes in the power-spectrum, already more work is needed, as discussed above. Finally, if the aim is to attach a well-defined physiological meaning to the observed CFC pattern, it is imperative to have either a generative model or additional external information, such as obtained from a direct perturbation of the putative physiological CFC mechanism, to link signal and underlying processes. Both of these latter approaches require a biophysical theory to be put forward as to how a neuron or an ensemble of neurons physically implements the coupling. For the moment, potential biological mechanisms of cross-frequency coupling are only starting to be discovered.

Indeed, whilst there is extensive knowledge about the physiological mechanisms responsible for different frequency components [1], not much is known about the cellular and network mechanisms of the *interactions* between these components [4]. Only recently some evidence about concrete mechanisms of interaction has been obtained from intervention studies in physiological systems and computational models. For example, by using transgenic mice it has been shown on the level of LFPs that hippocampal theta-gamma coupling depends on fast synaptic inhibition [31] and NMDA

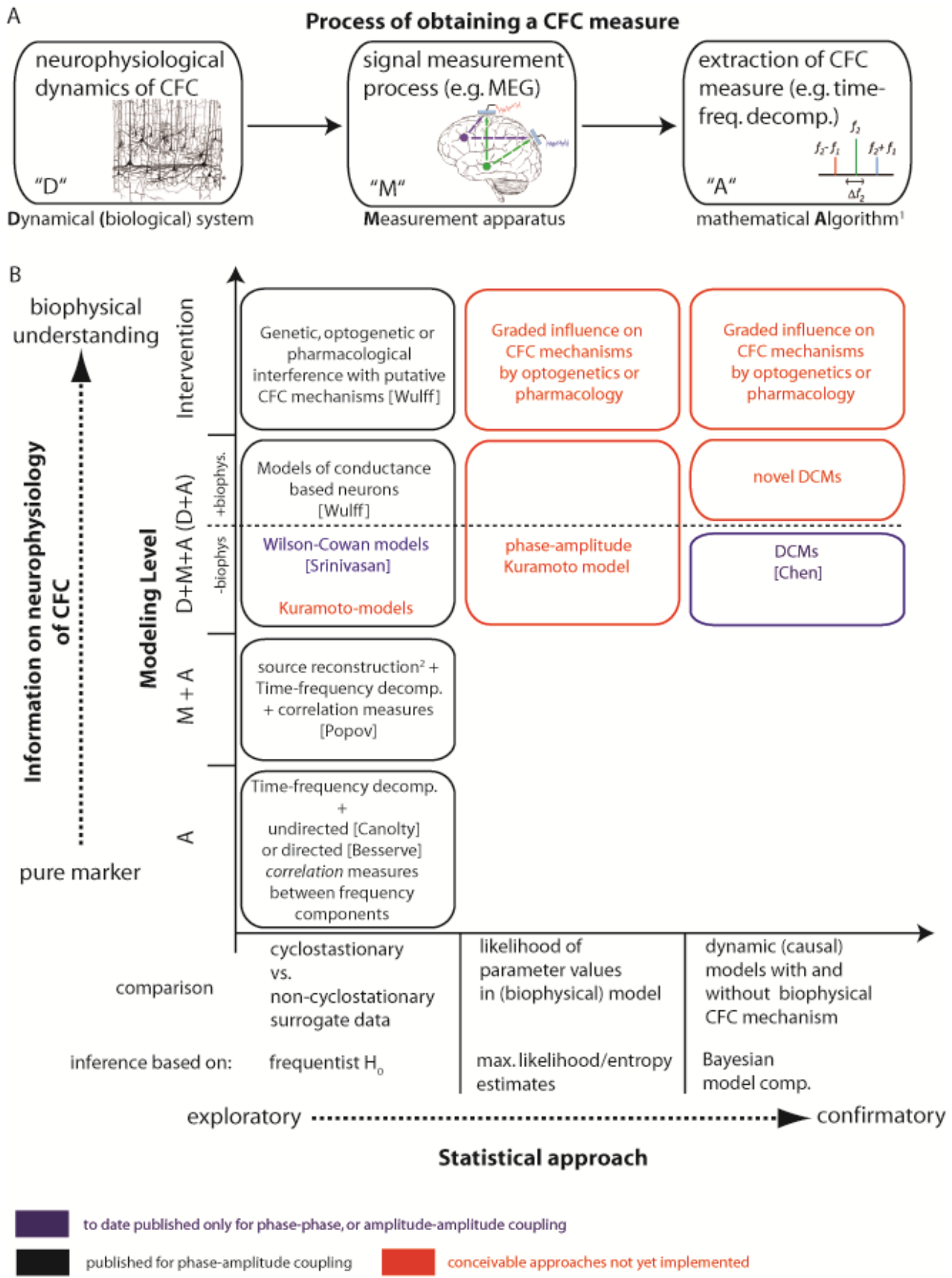

**Figure 4.**
The figure legend is on the next page.

**Figure 4 (continued).**
Organization of approaches to CFC. (A) The process by which a researcher obtains a CFC measure. "D" signifies the biological dynamical system (that may or may not have biological CFC). "M" signifies the physical transduction and measurement process, including all physical (unavoidable) filtering distortion, instrument noise and potentially mixing processes. The measurement process can generate CFC in absence of biological CFC. "A" signifies the mathematical algorithm applied to the measured data to obtain a figure of merit for CFC. Typically this step involves filtering or a time-frequency decomposition, and a linear or nonlinear correlation measure. As shown in the main text, also this process can give rise to CFC in absence of biological coupling, e.g. by ignoring the limits of time-frequency analysis in the face of non-stationarities. (B) Two-dimensional organization of CFC approaches. The x-axis sorts the approaches by the statistical inference technique that is used. Frequentist $H_0$ based approaches just test for presence of absence of CFC in the measured data, while maximum likelihood or Bayesian approaches perform inference on coupling parameters in models – and hence only appear for dynamic models with coupling parameters. The y-axis indicates the part(s) of the process in (A) that are modelled: "A" – just extracting a CFC measure from measured data is equivalent to modelling this extraction process itself. "M+A" – the generative model now comprises the measurement process and the measure extraction. "D+M+A" – the modelling comprises a dynamic model of the biological process, either via a dynamic proxy process that models only the bare essentials of the dynamics (like a Kuramoto model for phase-phase coupling), or has biological detail (like a Hodgkin-Huxley model with some explicit mechanisms implementing CFC). Approaches that were only published for phase-phase or amplitude-amplitude coupling are highlighted in blue text, conceivable approaches, that have to our knowledge not been implemented at all yet are highlighted in red.

($^1$) The fact that we use some numbers (measured data) and feed them to a mathematical algorithm to obtain other numbers (the CFC measure) can be modelled by just executing the algorithm. Nevertheless it is a part of the CFC process model.

($^2$) By source reconstruction we mean any inversion of the measurement process, e.g. unmixing via ICA, electromagnetic source reconstruction in electro- or magnetoencephalography, or removal of measurement noise.

We provided some representative references on the figure: Besserve, M., Scholkopf, B., Logothetis, N.K. & Panzeri, S. Causal relationships between frequency bands of extracellular signals in visual cortex revealed by an information theoretic analysis. *Journal of Computational Neuroscience* **29**, 547-566 (2010). Canolty, R.T.*, et al.* High gamma power is phase-locked to theta oscillations in human neocortex. *Science* **313**, 1626-1628 (2006). Chen, C.C.*, et al.* A dynamic causal model for evoked and induced responses. *Neuroimage* **59**, 340-348 (2012). Popov, T., Steffen, A., Weisz, N., Miller, G.A. & Rockstroh, B. Cross-frequency dynamics of neuromagnetic oscillatory activity: Two mechanisms of emotion regulation. *Psychophysiology* **49**, 1545-1557 (2012). Srinivasan, R. Thorpe, S. & Nunez, P.L. Top-down influences on local networks: basic theory with experimental implications. *Front Comp Neuroscience* **7**, 1:15 (2013). Wulff, P.*, et al.* Hippocampal theta rhythm and its coupling with gamma oscillations require fast inhibition onto parvalbumin-positive interneurons. *Proc Natl Acad Sci U S A* **106**, 3561-3566 (2009).

receptor-mediated excitation of parvalbumin-positive interneurons [32]. CFC at the LFP level has also been observed between alpha oscillations at the infragranular layer and gamma activity at the supragranular layers [33]. Recently, it was also shown that feedback inhibition enables CFC at the level of membrane potential fluctuations [34]. Thus, biological mechanisms for CFC can occur at the population level, at the single neuron level, or both. The most parsimonious explanation to account for these findings is that the low frequency oscillation reflects periodic fluctuations of the membrane potential and thus excitability, which in turn gate the occurrence of higher frequency activity in a phase specific manner [34]. Typically this higher frequency activity reflects spikes which could display a rhythmic pattern (possibly reflected as gamma oscillations).

Along with plausible biological mechanisms, one also needs to take into account possible alternative explanations for non-zero CFC measures. In this perspective we have tried to aid the interpretation of CFC by demonstrating that although CFC patterns are typically interpreted as reflecting physiological coupling, they can be also generated 1) by biological processes unrelated to direct coupling between different neural processes, and 2) by methodological pitfalls. In the following section we compiled some practical recommendations to help to avoid some of these errors. Taking these alternative explanations into account and controlling for them experimentally will help towards a clear interpretation of CFC results.

As laid out above, interventions would be ideal to study CFC. When an intervention is not possible, another principled approach to test for the presence of biophysical CFC is a formal comparison of computational models that do or do not incorporate biophysical CFC mechanisms with respect to their ability to explain the observed data. This could be done for example using sufficiently detailed Dynamical Causal Models and Bayesian model comparison [35]. If neither intervention nor formal model comparison are feasible, the researcher would have to limit the interpretation of observed CFC patterns (see Fig. 4) to that of a marker. This hierarchy of approaches is also reflected in the arrangement of methods in Figure 4.

To further exemplify the hierarchy of approaches illustrated in Figure 4, we turn to a different, more established measure - spectral power analysis. For example, measures of LFP power in the gamma band can simply be used as markers to classify different conditions (lower left corner in Fig. 4B). However, we also have several reasonable biophysical models (conceptual and computational) about the generating mechanisms of hippocampal and cortical gamma oscillations. These mechanisms were ultimately identified by interventional approaches (pharmacological, genetic, lesion) in a variety of physiological systems, as well as in computational models [36]. This means the field of spectral power analysis can draw on interventional approaches as well as formal model comparisons (upper row / right column of Fig. 4B). Nevertheless, we note that even for the mature field of spectral power analysis a change in LFP power in the gamma band can be due to several biophysical mechanisms (change in the number of neurons engaged in oscillations, their synchrony, etc) and these are not mutually exclusive. However, we can still map changes in gamma activity to a limited number of mechanistic options, each of them relatively well understood. Furthermore, since we are aware of several alternative explanations such as eye-movement artifacts that might contribute to gamma-band power changes we can design experiments to control for them [37]. We believe that similar steps will be required before CFC patterns in a signal can be confidently linked to any concrete biophysical mechanism.

## Practical recommendations

As previously discussed we will need progress in several directions to establish phase-amplitude or other types of CFC as a fundamental mechanism in coordinating neuronal activity. Together with experimental and modeling advances, stricter standards in the use of CFC metrics are also necessary. Below we list practical recommendations to avoid some of the mentioned caveats and increase the specificity of the most popular phase-amplitude CFC metric (see Fig. 1). Rather than a comprehensive algorithm this list should be thought as a check list that should help to minimize technical pitfalls and over-interpretation of phase-amplitude CFC measures in macroscopic signals.

*1    Presence of oscillations.* Signatures of oscillatory processes with clear peaks in a time-resolved power spectrum are indispensable prerequisites. The frequency component for defining the instantaneous phase should include one of the peaks.

*2    Selection of bandwidths.* The frequency band used to define the instantaneous phase should isolate energy associated with the oscillatory component of interest. If the center frequency is relatively stable a natural choice for the bandwidth can be directly obtained from the width of the corresponding peak in the power spectrum. The latter can be estimated by subtracting from the real power spectrum the power spectrum of baseline or a fit of the background power spectrum [38]. Note that the band defining the instantaneous amplitude at the higher frequency must be large enough to fit the sidebands caused by the assumed modulating lower frequency band (Figure 2A) and the lower frequency band should be narrow enough to define a meaningful phase. Therefore, adaptive rather than fixed bandwidths might be necessary when scanning the modulating frequency in explorative analyses.

*3    Interpretation of instantaneous phase.* A meaningful interpretation of instantaneous phase requires its monotonic growth in time. The presence of phase slips or reverses (also observed as negative instantaneous frequencies) must be checked and justified.

*4    Precision.* The precision of the method used to assign an instantaneous phase and amplitude to a signal should be determined for each analysis. The precision of the computation of the Hilbert transform of a signal *s(t)* can be estimated from the variance of *s(t)+H$^2$(s(t))*, which analytically should be identical to zero. Given the non-locality of Hilbert (or wavelet) transforms, edge effects can be severe. It is recommended to discard at least a few characteristic periods of the signal at the beginning and end of each segment of interest.

5    *Testing for non-linearities.* Non-linear responses to input or nonlinearities during the signal transduction can contribute to phase-amplitude CFC. The presence of harmonics in the signal should be tested by a bicoherence analysis and its contribution to CFC should be discussed. Partialization of phase-phase and amplitude-amplitude coupling is necessary to assess the role of non-linearities in generating spectral correlations (see *Supplementary results: Atmospheric noise shows CFC after squaring the signal; Small static non-linearity in ECoG data generates CFC; Mathematical example of a non-linearity; Supplementary discussion: The transitivity of correlation between phase and amplitude*).

6    *Testing for input-related non-stationarities.* When the timing of neuronal input to the recorded area is available, an analysis of relative locking between phase, amplitude and input can inform about the origin of the correlations.

7    *Temporal structure.* Information about the temporal structure of the putative interaction (e.g., sustained during many cycles versus a transient coupling) can be helpful to better characterize a presumed CFC and to disambiguate its origins. The modulation index only offers an average measure of CFC by computing the distance from a phase-amplitude histogram to a uniform distribution. However, such histogram can be used to identify the phase at which the average amplitude of high-frequency activity is maximal. The time series obtained by sampling the amplitudes of high-frequency activity at that particular phase can be used to provide some information about the temporal dynamics of the coupling.

8    *Surrogates.* Surrogate data should be created that minimally interfere with the phase and amplitude dynamics. For continuous recordings random point block-swapping is preferred over phase scrambling or cutting at several points.

9    *Specificity of effects.* Differences of CFC indices across conditions should be controlled for the differences in power at the presumed bands of interaction. The specific role of the coupling can be better assessed once the explanatory level of the power spectrum has been accounted for: When trial-based measures are available, stratification techniques should be used to compare subsets of trials for which the distributions of power at the bands of interest are identical across the conditions.

# Conclusions

Cross-frequency coupling (CFC) might be a key mechanism for the coordination of neural dynamics. Several independent research groups have observed cross-frequency coupling and related it to information processing, most notably to learning and memory [4-8]. Recently, CFC has also been used to investigate neurological and psychiatric disorders [9-15]. Thus, CFC analysis is potentially a promising approach to unravel brain function and some of their pathologies.

In the present manuscript we have reviewed some confounds that hamper phase-amplitude CFC analysis. Importantly, these confounds have not been considered in a significant percentage of recent publications and may have contributed to over-interpretations. This is a serious issue that needs to be resolved because CFC analysis is potentially a powerful tool to reveal fundamental features of neural computations. An obvious first step is to adopt stricter standards and canonical procedures for CFC analysis. To this end we suggested a – probably incomplete – list of controls that should be routinely checked. We have also attempted to organize the current modeling/statistical approaches to CFC in order to better identify their respective advantages and pitfalls and to point out where further methodological advances are required. We close by suggesting to always use the term "cross frequency correlation" instead of "coupling", unless coupling is unequivocally demonstrated.

# Acknowledgements

This work was supported by the Max Planck Society. RV, VP and MW were supported by LOEWE grant "Neuronale Koordination Forschunsgschwerpunkt Frankfurt (NeFF)". RV also acknowledges the support from the Hertie Foundation and the projects PUT438 and SF0180008s12 of the Estonian Research Council. VP received financial support from the German Ministry for Education and Research (BMBF) via the Bernstein Center for Computational Neuroscience (BCCN) Göttingen under Grant No. 01GQ1005B.

Supplementary materials for

# Untangling cross-frequency coupling in neuroscience

Juhan Aru, Jaan Aru, Viola Priesemann, Michael Wibral,

Luiz Lana, Gordon Pipa, Wolf Singer, Raul Vicente

**This PDF file includes:**

*Materials and Methods*

### Analysis of CFC

All analyses were performed using custom built MATLAB routines. To extract the frequency components, all signals were band-filtered with a two-way least-squares finite-impulse-response (FIR) filter (eegfilt.m from the EEGLAB toolbox [1]). For each component of interest, instantaneous phase and amplitude were estimated by the analytical signal approach. The Hilbert transform was applied to the filtered signals in order to define the imaginary part of the complex-valued time series (with real part being the filtered signal). The polar coordinates of these analytical signals define the instantaneous amplitudes and phases.

The time-dependent power locked to phase-troughs of slow components used to create Supplementary Figures 1B, 2C, 3E, and 6 was extracted as described in the supporting information of Ref.[2].

The modulation index used in Supplementary Figure 2 (D), and Supplementary Figure 6 (bottom panels) followed the original formulation by Tort et al. [3]. All modulation indices and histograms of mean amplitudes were computed for 20 equally-sized bins for the phase variable.

### Data collection for Supplementary Figure 3

We recorded neuronal activity from the retina and the tectum of the turtle (Pseudemys scripta elegans) to analyze the effect of stimulus input on cross-frequency coupling.

*Preparation.* Experiments were approved by the German local authorities (Regierungspraesidium, Hessen, Darmstadt). One turtle (Pseudemys scripta elegans) was anesthetized with 15 mg Ketamine, and 2 mg Medetomidinhydrochloride and decapitated. The entire brain with the eyes attached was removed as described in Ref. [4]. The brain was placed in a petri dish and superfused with oxygenated ringer. The ringer consisted of (in mM) 96.5 NaCl, 2.6 KCl, 2.0 $MgCl_2$, 31.5 $NaHCO^3$, 20 D-glucose, 4 $CaCl_2$ at pH 7.4 and was administered at room temperature (22 C).

*Electrophysiological recordings.* The electroretinogram was recorded with a chlorided silver wire in a Vaseline well that was built around the right eye. The tectal signal was recorded in a superficial layer at the center of the left tectum with a quartz/platinum-tungsten electrode (Thomas Recordings, Giessen, Germany) with impedance 1 MΩ at 1 kHz. Data were amplified and filtered (1 Hz to 6 kHz) before being digitized at 32 kHz. For the analysis, data were low-pass filtered with 240 Hz, down sampled to 500 Hz and cut into 60 trials with 50 s each.

*Visual stimulation.* A sequence of red LED light pulses with random duration (uniform distribution between 1 ms and 2 s) and random inter pulse interval (uniform distribution between 1 ms and 5 s) was triggered via the parallel port using MATLAB and the Psychophysics Toolbox extension [5,6]. A light guide projected the full field flashes onto the retina.

### **Data collection for the Supplementary Figure 4**

*Recordings.* We analyzed electrocorticograms from 2 subjects with pharmacoresistant epilepsy who had implanted strip electrodes (AD-Tech) on their visual cortex for diagnostic purposes. The electrodes were referenced to linked mastoids, amplified (Schwarzer GmbH), and recorded at a sampling rate of 1000 Hz. The location of electrode contacts was ascertained by MRI.

*Visual stimulation.* The subjects were presented with brief (150 ms) noisy images with or without a person in them. The subjects had to indicate via a button press whether they had perceived a person or not and whether the person was male or female. The pre-stimulus time window ranged from -1000 ms to 0 relative to stimulus onset. The post-stimulus time window ranged from 0 to 1000 ms. The response screen appeared only after this window, thus our analysis windows did not contain motor responses. Supplementary Figure 4 shows the results of the CFC analysis for one of the representative electrodes in the vicinity of the extrastriate body area (MNI coordinates -56, -67, -8).

*Supplementary Results*

**<u>Literature review</u>**

To assess the prevalence of the critical issues presented in the current manuscript we conducted a literature review. In particular, we evaluated publications that appeared in the recent years (2010 - 2013 and January 2014) to demonstrate that the caveats discussed in the main text are timely. To show that these caveats are ignored even in the key journals in neuroscience, we focused our analysis on publications from Science, PNAS, Neuron, Journal of Neuroscience and Neuroimage (our search terms did not yield papers in Nature or Nature Neuroscience). We searched for articles in PubMed with terms "cross-frequency coupling", "phase-amplitude coupling", "cross frequency interactions" or "nested phase amplitude". We added manually one high-ranking paper which was missed by these searches. We excluded three papers that were returned, but focused on phase-phase or amplitude-amplitude coupling. In total, 22 articles were evaluated according to five criteria, each related to one of the main issues discussed in the main text:

- *Phase interpretability.* As stated in the main manuscript and the supplementary discussion, a meaningful interpretation of a phase variable, and thus of a phase-amplitude CFC measure, requires a clear peak in the power spectrum. Therefore, we first assessed whether a publication provided evidence for a spectral peak for the modulating component's frequency. We concluded that the publications did if there either was a remark about a respective spectral peak in the main text or in the supplementary material or if any of the figures presented either a power spectrum or a time-frequency representation which allowed us to conclude that there was a spectral peak. Else, we categorized the publication as not providing evidence for a spectral peak. We also categorized the publication as not providing evidence for a spectral peak if more than 50 different channels were analyzed and it was not stated in the text that the authors took care that each channel which had or was part of a significant coupling had a spectral peak for the modulating frequency.

- *Selection of bandwidth.* We studied how the bandwidth selection or scanning was done. This was usually straightforwardly reported in the methods section. In detail, we evaluated whether the bandwidth of the high-frequency component included the sidebands induced by the modulating (low) frequency component (see Figure 2). If the authors just scanned the modulated (high) frequency range with some fixed frequency bandwidth that did not always include the sidebands, the paper was classified as not providing a justification for the chosen bandwidth.

- *Non-stationarity.* As explained in the main text, non-stationary input to a given neural population can create significant CFC estimates even if there is no physiological coupling. We assessed whether the authors discussed the effects of non-stationary inputs as potentially leading to the observed CFC estimate.

- *Surrogates.* We considered how the surrogate distribution was obtained. As we reported in the main text, the most conservative surrogate data from block resampling approaches are obtained by minimizing non-stationarities introduced at cut-points. For continuous recordings, this is achieved via cutting only at a single point at a random location. For event-related data, this is achieved via using trial-shuffling. If surrogate data were constructed differently, e.g. if phase scrambling was used or if continuous data was cut in many points, we classified the paper as not constructing the most conservative surrogates. If the authors did not use a surrogate distribution, e.g. when the effect was computed by comparing two conditions, we did not count the publication. The computation of the surrogates was usually reported in the methods section.

- *Control for spectral changes across conditions*. We took into account only papers that compared CFC between different conditions and reported differences. As explained in the main text, differences in spectral power can lead to differences in CFC without a true change in coupling strength. We assessed whether the effects of potential differences in spectral power were taken into account or controlled for. We classified the paper as not having controlled for the effects of differences in power spectra if such differences were reported, but not addressed. We classified the paper as taking care of the differences in the power spectrum if the authors took steps to deal with the issue.

**Results of the literature review**

| Question | Yes | No | % Yes | Papers not counted |
|---|---|---|---|---|
| Are spectral peaks identified for the (modulating) low-frequency component? | 10 | 12 | 45.45% | |
| Is there justification for the chosen bands? | 3 | 19 | 13.36% | |
| Is the possibility of non-stationary input leading to observed CFC patterns discussed? | 3 | 19 | 13.36% | |
| Are the surrogates constructed in the most conservative way? | 10 | 4 | 71.43% | 8 |
| Are the differences in the power spectrum accounted for? | 8 | 7 | 53.33% | 7 |

## **Examples of spurious cross-frequency coupling (CFC)**

Cross-frequency coupling analysis is aimed at detecting specific spectral correlations. However, the origins of such correlations are diverse and not always reflect interactions across frequencies. Here we first show that a single oscillator can readily exhibit cross-frequency coupling features simply by virtue of its non-linear properties. We illustrate this case with the well-known Van der Pol system which is a generic model of non-linear relaxation oscillators. Its evolution is given by the following differential equation

$$\frac{d^2 x}{dt^2} - \mu(1-x^2)\frac{dx}{dt} + x = 0,$$

where $\mu$ represents a non-linear damping coefficient. Supplementary Figure 1A shows the time series of the oscillator for $\mu = 3$. Supplementary Figure 1B shows the power of high-frequency components locked to the trough of the phase of a low frequency band around the fundamental frequency of the oscillation (band: 0.005 - 0.02 cycles/time unit). Clearly, a phase dependence of high-frequency power is observed. However, such systems are not necessarily decomposable into two subsystems oscillating at different frequencies and causally interacting. Instead, the spectral correlations can be simply related to the shape of the oscillatory orbit and the non-linear characteristics of a single oscillator. In particular, the non-linear damping (damping dependent on the state of the oscillator) is responsible for an increasing slope for certain states of the oscillator which is reflected in high-frequency power being associated with certain phases of the fundamental oscillatory frequency.

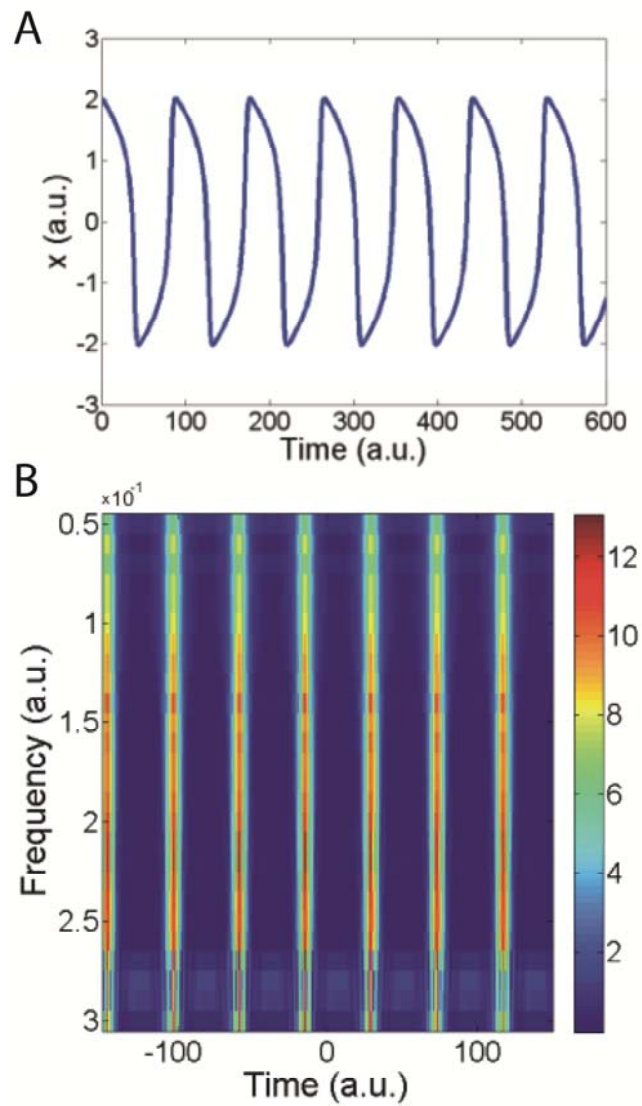

**Supplementary Figure 1.** Cross-frequency coupling can arise in a single non-linear oscillator. **A.** time series of a Van der Pol oscillator with non-linear dumping coefficient $\mu$ = 3. **B.** power locked to the trough of the phase of a low-frequency component (low frequency band (0.005-0.02 cycles/time unit)). Color coded is the power of the high-frequency component. CFC analysis can confound non-linear features of oscillations as interaction across frequencies.

Another example of how non-zero CFC estimates do not imply interactions across frequencies is given by a simulated spiking process. For the time series we have taken numbers 1, 2 ... 30000 and set them to 1 if the number was prime and 0 otherwise. The sampling rate of this signal was nominally set to 1000 Hz. With this procedure we constructed a point process in which spikes occur at the bins that are prime numbers. As CFC analysis is performed for continuous signals and not for point processes as in the example above, we convolved our time series with a smooth kernel (an alpha-function with a time constant of 5 ms mimicking the conductance response of a synapse to incoming spikes). As evidenced in Supplementary Figure 2 and confirmed by calculating the CFC measures, the artificial series also shows high CFC estimates (using code from [3]). In this example, the spiking events anchor both the phase of slow frequency bands and the high-frequency components of the kernel function, which automatically leads to CFC.

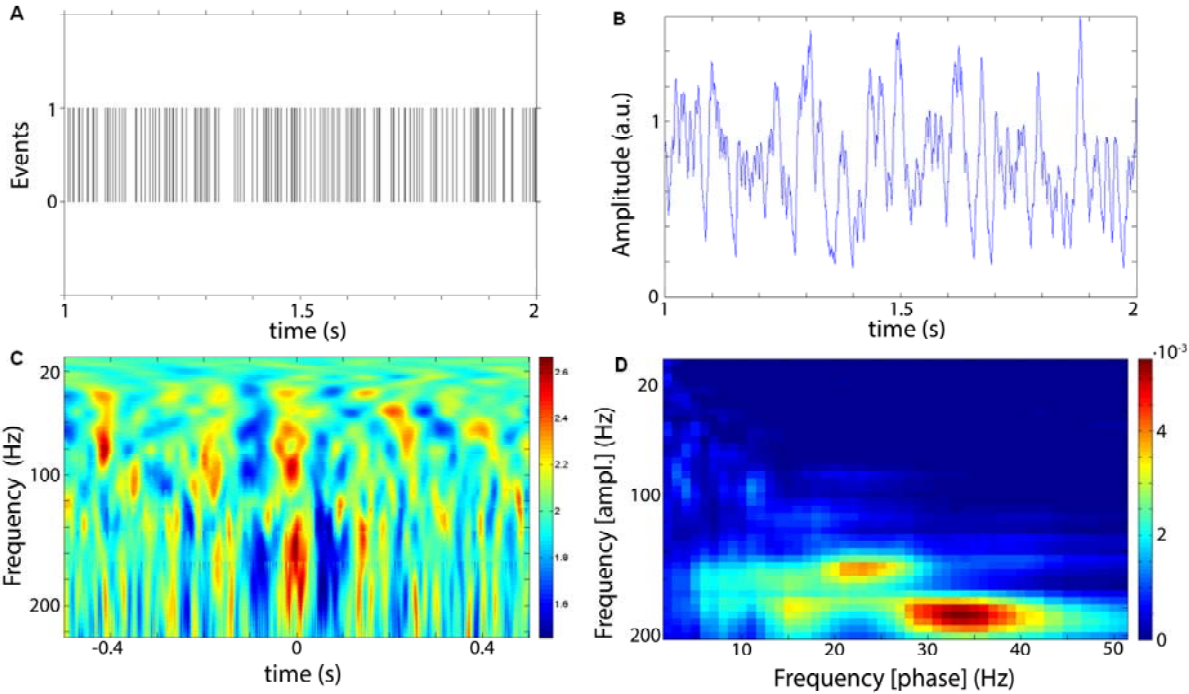

**Supplementary Figure 2.** Point processes (even when convolved with a smooth kernel) lead to CFC. **A.** Point process where spiking events occur at bins that are prime numbers. **B.** Convolution of this point process with an alpha function. **C.** Power of high-frequency activity locked to the trough of low-frequency activity (4-8 Hz). **D.** Kullback-Leibler divergence between the amplitude-phase distribution of the time series and the uniform distribution ("modulation index" from Tort et al. [3]).

Another curious time series for which we found CFC effects are the intervals between consecutive zeros of the Riemann Zeta function (only zeros on the axis with real part ½ were considered). Results are not shown.

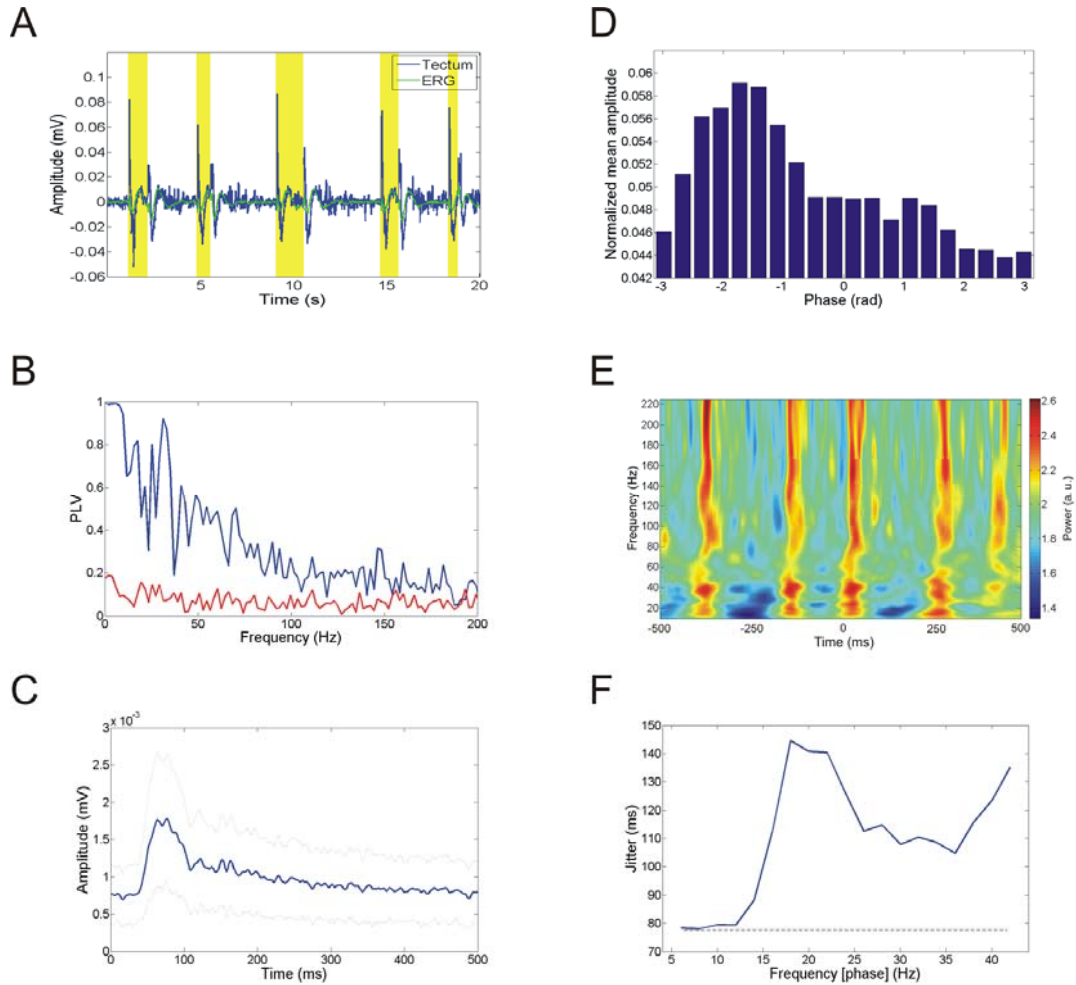

**Supplementary Figure 3. A**. Time series of stimulus and recorded activity in the retina and optic tectum of the turtle. The rest of the panels are based on the electroretinogram only but similar results hold for the optic tectum. **B**. Phase locking values (PLV) as a function of frequency for a period following (blue) or preceding (red) the onset of visual stimulation. **C**. Amplitude of high-frequency activity (80-200 Hz) locked to stimulus onset (mean is represented in blue while mean plus/minus standard deviation are in grey). **D**. Amplitude-phase histogram for the whole time series (100 seconds). **E**. Power locked to the trough of a low-frequency band (4-8 Hz). **F**. Jitter from the peak of high-frequency (80-200 Hz) power to stimulus onset (71 ms; grey line) is always smaller than to the trough of the phase of lower frequency bands closest to their maximal power (blue line). This means that most likely input simultaneously affected both the phase of the low-frequency component and the amplitude of the high-frequency component, thus generating a significant CFC estimate without underlying physiological coupling.

### Effect of non-stationary input on CFC

As mentioned in the main text, phase-amplitude coupling measured anywhere in the brain can be potentially explained by common influence on the phase and amplitude, without the phase of a low frequency oscillation modulating the power of high frequency activity.

To illustrate this point we analyzed the CFC patterns of the electroretinogram (ERG) and LFP recorded in the optic tectum of a turtle (Pseudemys scripta elegans) during visual stimulation (Supplementary Figure 3 and *Supplementary methods*). The comparison of phase locking values for windows placed before and after the stimulus onset revealed an increase of phase locking preferentially for lower frequencies (Supplementary Figure 3B). At the same time, the stimulus elicited an increase in power, including the power of high-frequency bands, 80-200 Hz (Supplementary Figure 3C). The combination of increased activity at high-frequencies and phase-locking to the stimulus of lower frequencies is sufficient to obtain significant measures in standard CFC analysis (Supplementary Figure 3D-E). As already mentioned in the main text, the evidence presented in Supplementary Figure 3D-E would suggest significant CFC. However, we can gain more insight into this issue by using information about the stimulus timing and studying how precisely the high-frequency component locks to both the stimulus onset and the phase of the low-frequency component. If the high-frequency amplitude was mainly coupled to the phase of slower oscillations, one would expect that the locking between these frequency components would be more precise than the one between the high-frequency component and the stimulus onset. We evaluated this locking precision by comparing the jitter between the high-frequency component and stimulus onset with the jitter between the high-frequency component and the low-frequency component. More precisely, this jitter was computed as the variance over trials of the difference between the time of maximal high frequency amplitude and stimulus onset, and between the time of maximal high-frequency amplitude and any phase of the lower frequency component, respectively. For our recorded data it turned out that the jitter between the high-frequency component and stimulus onset was always smaller than the jitter between the high-frequency component and any phase of the low-frequency component (Supplementary Figure 3F). This suggests that the high-frequency component is more affected by the stimulus than by the phase of any low-frequency component.

Next we analyzed CFC in human intracranial recordings. As in the turtle experiment, the brief visual stimulation led to an overall increase of power (Supplementary Figure 4A, C). Associated with the stimulus, phase was reorganized predominantly at low frequencies, as measured by phase locking value (Supplementary Figure 4B). As a consequence of stimulus-related power and phase adjustments, the mean amplitude-phase histogram following stimulation is non-uniform (Supplementary Figure 4D). The pairing amplitude and phase of different trials renders the histogram uniform (Supplementary Figure 4E). To gain insight on whether the observed cross-frequency coupling is due to an interaction across frequencies or simply an effect of the stimulus, a jitter analysis is presented in panel F. Jitter (as measured by the standard deviation of a series of time differences) from the peak of high-frequency power (80-200 Hz) to stimulus onset is around 80 ms (grey line). The black line represents the jitter from the peak of high-frequency power to the trough of the phase of the slow components closest to their maximal power. For a range of frequencies, the jitter between power and stimulus onset is smaller (more locked) than the one between power and phase (less locked). This suggests that a common drive from the time-varying input is another plausible explanation for the correlation between frequency components, in addition to a putative interaction.

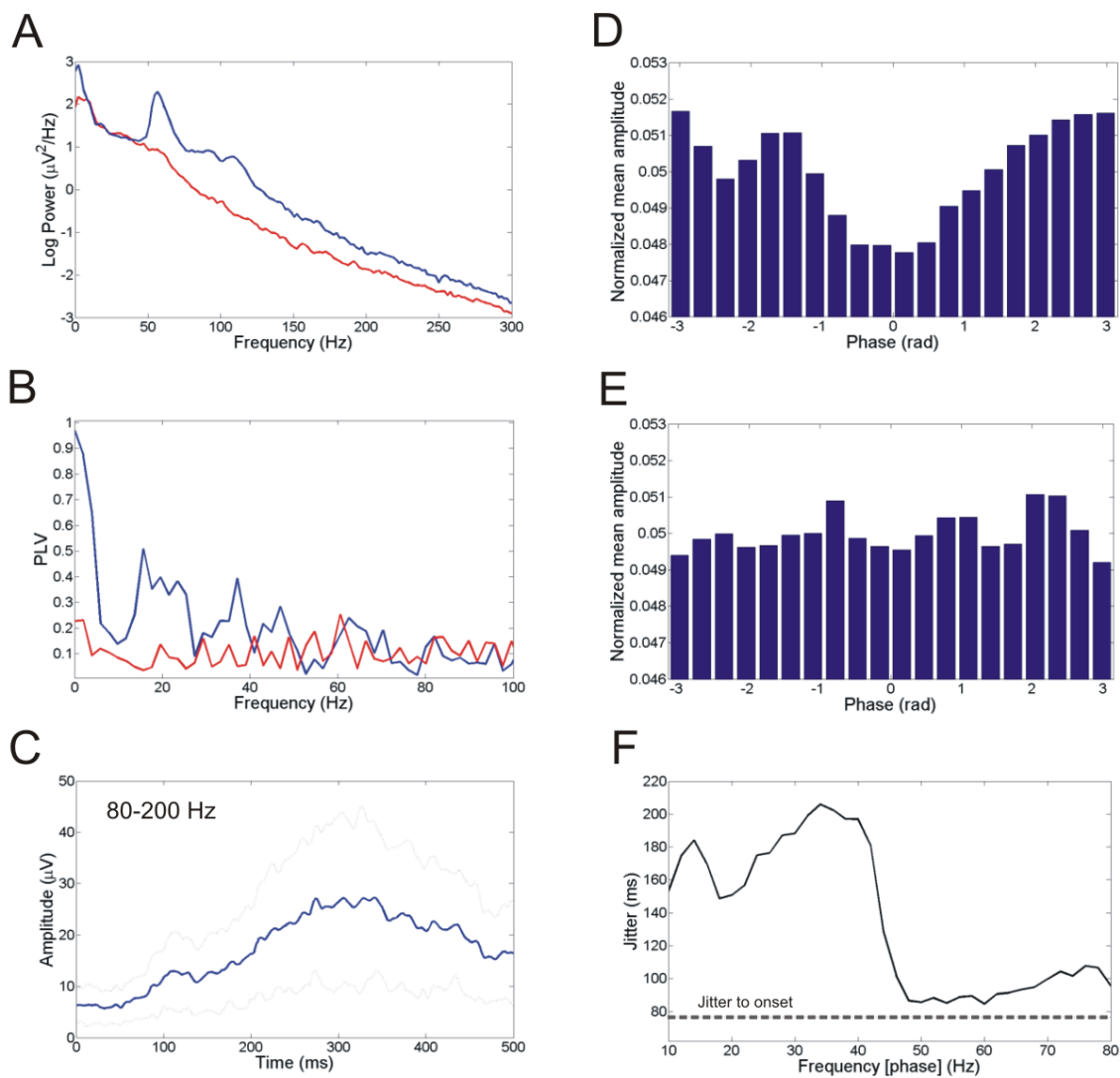

**Supplementary Figure 4.** Analysis of human ECoG in a high order visual area showing that non-stationary input leads to positive measures of CFC. **A.** Power Spectra for a period following (blue) or preceding (red) the onset of visual stimulation. **B.** Phase locking values as a function of frequency (colors as in A). **C.** Amplitude of high-frequency activity (80-200 Hz) locked to stimulus onset (mean is represented in blue while mean plus/minus standard deviations are in grey). **D.** Amplitude-phase histogram for the time window of 500 ms following stimulation. **E.** Same as D, but pairing amplitude and phase from different trials. **F**. The jitter from the high-frequency (80-200 Hz) power peaks to the stimulus onset (77 ms; grey line) is always smaller than the jitter measured from the lower frequency trough closest to the maximal high frequency power (blue line).

## **Phase-amplitude coupling for event-related potentials (ERPAC)**

Recently, a new approach [7] has been developed to measure transient phase-amplitude coupling directly in an event-related manner. This approach, called ERPAC, was designed to overcome the problem of spurious CFC that is observed due to event-related non-stationarities in the signal, i.e. event driven changes in phases and amplitudes (see the turtle data example above). ERPAC is actually one algorithm in a large class that exploits the fact that locally non-stationary processes can be made quasi-stationary if they are repeatable in some sense (such as trials in an ERP experiment). Algorithms in this class create stationarity via sampling features only from fixed time points in a repetition (see Ref. [23] for details on achieving stationarity by targeted sampling). For the method in [7] these features are the Hilbert amplitudes and phases, and the fixed time points are taken with respect to the timing of an experimental manipulation, e.g. a stimulus. At first sight this method indeed seems to solve the problem of spurious CFC due to input-related non-stationarities. However, as we will show below, this holds only for the case of very precisely repeated responses. This is because only for very precisely repeated responses ('additive evoked components' in Ref. [24]), the desired stationarity over trials is obtained. The condition of precisely repeated responses is at most met in subcortical and cortical input stages of sensory systems. Elsewhere it is not met, despite the presence of a clear ERP [24]. If there is no precisely repeated response, i.e. if there is a slight jitter in input arrival times to a brain area over trials, then the known problems of non-stationarities - that lead to spurious CFC - arise again.

To show this in an example, we have simulated a small jitter in arrival times of an input that leads to both a phase reset of low frequencies and an amplitude increase in high frequencies (Supplementary Figure 5C). In the numerical simulations shown in Supplementary Figure 5 the considered signal is composed by white noise plus two sinusoidal components (6 Hz and 50 Hz) that were modulated by a Gaussian profile (mean = 1320 ms; standard deviation = 100 ms) whose center is randomly jittered over trials (jitter = 40 ms). ERPAC measures were obtained from the code available at http://darb.ketyov.com/professional/publications/erpac.zip. As expected from the argument above, for the case where the input response (Gaussian profile) modulates both the amplitude of the high-frequency component and resets the phase of the low-frequency components, the method from [7] detects spurious CFC (p-values < $10^{-11}$) that is purely input driven. We note that this is ultimately a variant of the problem of unknown timing of input arrival presented in Figure 3 of the main text.

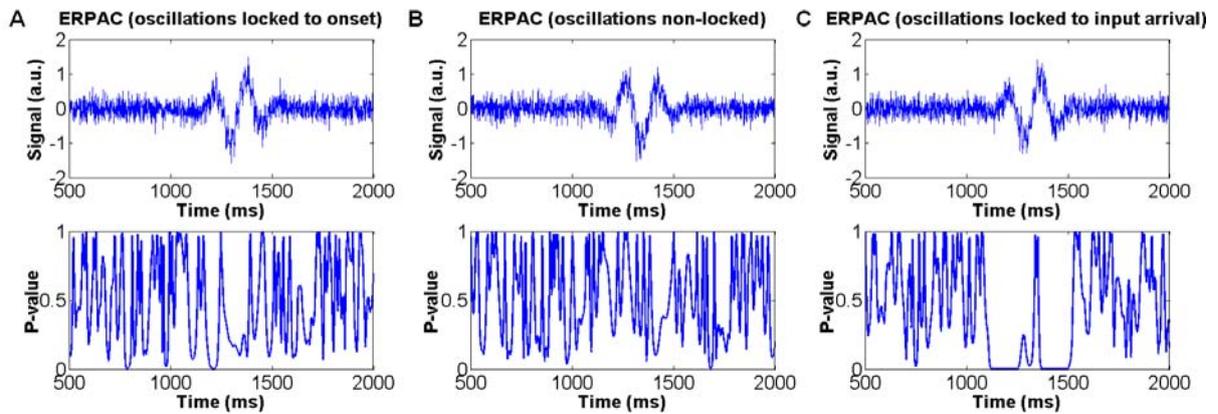

**Supplementary Figure 5.** Event-related phase-amplitude coupling (ERPAC) as detected by the method of Voytek et al. [7] **A.** One representative trial (top) and ERPAC (bottom) measure for a signal with oscillatory components phase locked to a fixed onset (at 1000 ms). **B.** One representative trial and ERPAC measure for a signal with oscillatory components not phase locked over trials. **C.** One representative trial and ERPAC measure for a signal with oscillatory components phase locked to a jittered input response (simulated by a Gaussian amplitude profile).

We also observed that in the concrete example above (Figure 5), as expected, already the *amplitude* of low frequency component explains some variance of the amplitude of the high frequency component. It is thus plausible and possible that replacing the bivariate generalized linear model used in [7] with a multivariate model (for example including the amplitude of the low frequency component), could more specifically delineate the various dependencies between frequency components.

## Atmospheric noise shows CFC after squaring the signal

Another concern with interpreting cross-frequency interactions is the role of non-linearities in creating such correlations. The signal provided by a neurophysiological recording can be influenced by different non-linearities. Some of them are intrinsic to basic neuronal processing such as action potential generation or nonlinear dendritic summation caused by voltage-gated conductances.

Others, however, might be of a more mundane origin. These non-linearities can include the electrical properties of the neuronal tissue (e.g., activity-dependent resistivity changes [8]) or small deviations from linearity occurring at any stage of the transduction from the neuronal generators to the final output signal that is subject to analysis.

To test the effect of static non-linearities in CFC analyses, we compared the CFC patterns of a random time series and its square. The random time series is composed of uniformly distributed random numbers taken from a physical source (10000 samples of atmospheric noise were taken from www.random.org). As shown in Supplementary Figure 6, random noise does not contain any evident structure of phase-amplitude coupling. However, the square of this random signal displays a clear modulation. Therefore, non-linearities can confound the CFC measures.

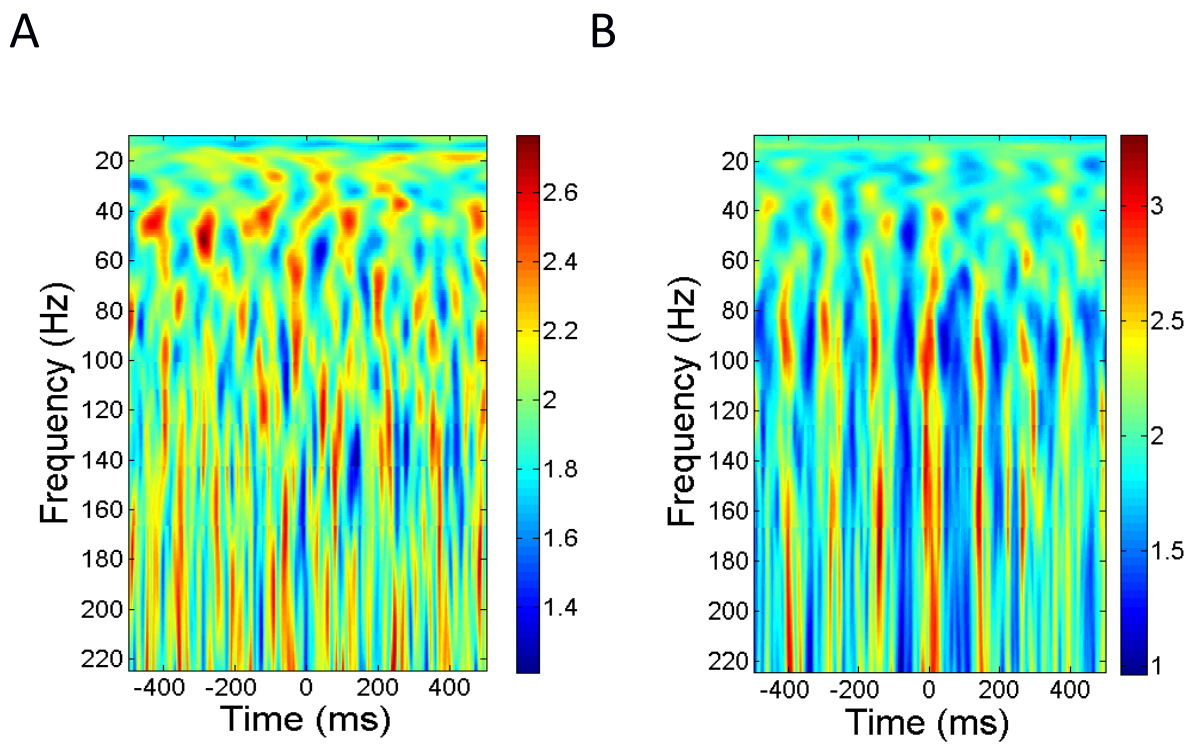

**Supplementary Figure 6.** Non-linearities can generate specific patterns of CFC even for random data (atmospheric noise, 10000 samples, nominal sampling rate 1000 Hz). **A.** Power at different frequencies locked to the trough of phase of a low frequency component (4-8 Hz). **B.** Same analysis as in top panel but for the signal obtained by point-wise squaring the atmospheric noise time series.

## Small static non-linearity in ECoG data generates CFC

Adding a statistically negligible contribution (Pearson coefficient of correlation between original and distorted signal ~0.99) of a squared signal to an ECoG signal not showing any significant CFC, will yield strong CFC (Supplementary Figure 7).

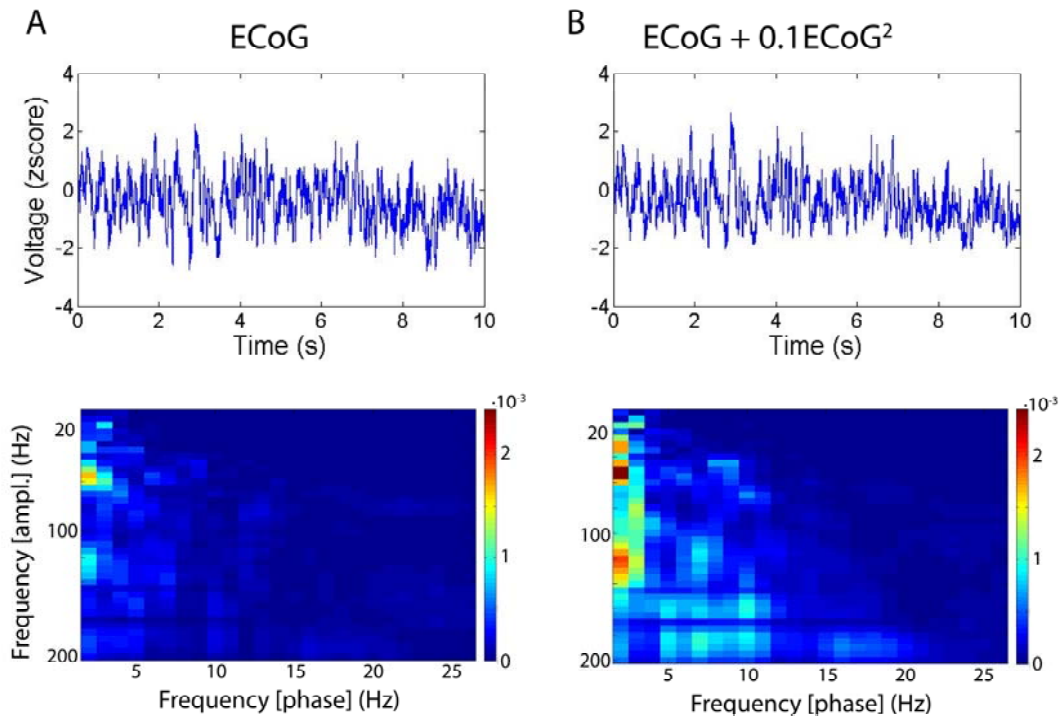

**Supplementary Figure 7. A.** Original ECoG time series (top) and modulation index (bottom). **B.** The same signal after the addition of 10% of its square, and the corresponding modulation index. Notice that the signals (top panels) look very similar, but the modulation index on the right is much stronger. This result shows that even small non-linearities can create spurious CFC patterns.

This result is natural as even small non-linearities readily create harmonics which can spread over the whole spectrum and, thus, generate long-distance spectral correlations for a continuous and large range of frequencies. Importantly, such non-linearities can create specific CFC patterns, where the greatest amplitude of the high-frequency component is related to a specific phase of the low frequency component. For a simple mathematical example of a non-linearity see next section.

## Mathematical example of a non-linearity

For a simple mathematical example of a non-linearity consider a harmonic wave *x(t)* with frequency *f* given by:

$$x(t) = \cos(2\pi ft)$$

and a non-linear version of *x* that to a second-order approximation is given by:

$$y = x + ax^2 \approx \cos(2\pi ft) + \frac{a}{2}\cos(2\pi(2f)t) + \frac{a}{2},$$

with the approximation holding if *a* << 1, i.e. if the non-linear contribution is small. In the following we show that the Hilbert amplitude of *y* is a function of the Hilbert phase of *x*, i.e. there is a clear mathematical dependence between the phase of *x* and the amplitude of *y*. To extract the phase component, we filter *y* only narrowly around the frequency *f* (thereby isolating *x*). To extract the amplitude component, we filter *y* widely (with a bandwidth of at least +/- *f* so that the side bands are included as required) around the frequency *2f* (thereby retrieving *y*). CFC analysis between *x* and *y* will then result in perfect phase-amplitude coupling, as follows.

The Hilbert phase of *x* is by definition seen to be *2πft* modulus *2π*. The Hilbert amplitude of *y* can be found by constructing the analytic signal of *y* by just using two facts: 1) the analytic signal of a harmonic oscillation is given by its complex form, and 2) Hilbert transform is linear. Hence, the analytical signal of *y* is:

$$Y(t) = \exp(i2\pi ft) + \frac{a}{2}\exp(i2\pi(2f)t) + \frac{a}{2}.$$

Now this can be rewritten as:

$$Y(t) = \exp(i2\pi ft) + \frac{a}{2}\left(\exp(i2\pi(2f)t) + 1\right) =$$

$$\exp(i2\pi ft) + \frac{a}{2}\exp(i2\pi ft)\left(\exp(i2\pi ft) + \exp(-i2\pi ft)\right) =$$

$$\exp(i2\pi ft)(1 + a \cdot \cos(2\pi ft)).$$

From here we can see that the Hilbert amplitude of *y* is equal to $1 + a \cdot \cos(2\pi ft)$

and hence, mathematically, the amplitude of y (the high frequency component) is a function of the phase of x (the low frequency component).

*Supplementary Discussion*

**Conditions for a meaningful phase**

Intuitively, a phase is a parameter that tells us where we are within a cycle of some repetitive motion or variation. The amplitude informs us about the span of that motion. For a simple harmonic motion ($s(t) = a*cos(\omega t)$), the amplitude *a* is a constant and the phase indexes the position along the cycle. Importantly, the phase is useful because it is an index that grows monotonically within a cycle, i.e., the larger the phase the larger the completed fraction of the current cycle. Geometrically, amplitude and phase correspond to the distance to origin and the angle subtended by the complex function $a*cos(\omega t)+i\, a*sin(\omega t)$, respectively. Notice that the imaginary part is a 90 degrees shifted copy of the original signal (cosine becomes sine).

When faced with a less regular signal *s(t)*, the task of assigning a phase and amplitude is more delicate since there are infinitely many pairs of functions $a(t)>0$ and $\phi(t)$ that satisfy $s(t) = a(t)*cos(\phi(t))$. Which one to choose? Gabor elegantly proposed the analytical signal approach to provide a unique and unambiguous solution [9]. The idea is the following: since we know how to define a phase and amplitude for a single sinusoidal, we can decompose the signal in its Fourier components and repeat a similar procedure for each component. This leads to the complex analytical signal. The real part of the signal is the original signal *s(t)* and the imaginary part is formed by 90 degrees shifted copies of the original signal's Fourier components. The Hilbert transform (*H*) sums up these decompositions and shifting operations, and the analytical signal of *s(t)* is thus represented as $s_a(t)= s(t)+i\, H(s(t))$.

However, as several authors have noted [10], a clear interpretation of the phase and amplitude obtained by this or, indeed, any other method is restricted to *narrow-band* signals. Thus, although the analytical approach yields an instantaneous phase and amplitude for any signal, the interpretation is only clear if the signal does not deviate much from being a smooth periodic function. Fortunately, the narrow-band condition covers the presence of moderate noise and smooth frequency fluctuations since in that case the signal will be still narrow-band but with a slowly varying center frequency. The main issue is that for narrow-band signals amplitude and phase can be considered as separate and independent entities, while this is not the case for broad-band signals. Take the example of a modulated signal $s(t) = a(t)*cos(\phi(t))$ in which some physical meaning is attached to *a(t)* and *ϕ(t)*. When *ϕ(t)* is a smooth function of time and *a(t)* changes slow enough within each cycle (i.e., an increase of *ϕ(t)* by $2\pi$), then *s(t)* will result in a narrow-band signal where phase and amplitude are separable variables. In this case the estimation of phase and amplitude via analytical signal analysis or other approaches can recover the exact modulated values of $\phi(t)$ and $a(t)$ from only $s(t)$. The technical reason for this result is known as the Bedrosian theorem [11]. This theorem also implies that for broad-band modulated signals where the spectra of $a(t)$ and $cos(\phi(t))$ overlap, the phase and amplitude estimated from the analytical signal will be composed by an intricate mixture of $a(t)$ and $\phi(t)$. Moreover, values of the estimated instantaneous frequency (seen as a derivative of the instantaneous phase) could even take negative values. Importantly, this is not a problem of any particular approach but it reflects the fact that irregular signals might have degrees of freedom that cannot be faithfully represented by just a smoothly changing phase-amplitude pair. Indeed, the analytical signal approach can be considered as the optimal two-dimensional delay embedding of a univariate signal to estimate its phase and amplitude. However, many signals need a higher number of dimensions to be fully unfolded [12]. In that case, insisting on the two dimensional description of the signal (by phase and amplitude or any other pair) results in projecting all the degrees of freedom into just two variables and thereby compromising their interpretability.

It is noteworthy then that phase-amplitude CFC is frequently investigated for regions in the spectrum that locally exhibit a power law, i.e., $P(f) = 1/f^{\alpha}$, and thus do not contain a peak [13]. For these regions, applying a narrow-band filter of a few Hz will render a signal with smooth phase and amplitude dynamics. However, this apparent smoothness is a simple result of filtering, and not a sign that there are processes in this spectral range that indeed have a smooth, or even meaningful, phase. Hence, narrow band filtering in regions exhibiting $1/f^{\alpha}$ power spectrum does not lead to an interpretable phase, despite its apparent smoothness. Thus, findings of spectral correlations at regions exhibiting $1/f^{\alpha}$ decays should be interpreted extremely carefully.

In sum, only a natural concentration of power around some center frequency in a time-frequency decomposition enables a meaningful interpretation of the phase, and, thereby, CFC analysis.

## **The importance of the bandwidth**

The concentration of power around some frequency is a necessary prerequisite for a meaningful interpretation of the phase of a signal. Thus, for the phase variable the center frequency and bandwidth selected should include a peak of the spectrum. Even under such condition the choice of the bandwidth severely affects the results. A too narrow filter, at least compared to the natural width of the peak, will result in a smooth and well-behaved phase which, however, is hardly representative of the underlying oscillatory but variable signal. A too broad filter can result in an ill-behaved phase with phase slips or reversals by incorporating 1/f components. Importantly, both cases can lead to a loss of sensitivity and interpretability of the analysis. A possible strategy is to precisely exploit the freedom of the bandwidth to look for a sweet point where the phase dynamics shows maximal robustness against small bandwidth changes. Another heuristic but natural choice is to select the bandwidth as the range around the peak that clearly stands out from the background of the power spectrum. If a good fitting for the background spectrum is available, then subtracting the real and interpolated spectra can deliver an estimate of the width [14]. Finally, if additional physiological information is available it might be possible to use functional criteria to select bands. For example, if interested in CFC of human occipital alpha rhythms it is probably more appropriate to find subject specific alpha bands rather than to select a fixed band ranging from 8 to 12 Hz.

Now we turn to the issue of selecting an adequate bandwidth for the frequency component defining the instantaneous amplitude. Following the example in Figure 2, suppose it is known that the lower frequency component is around $f_1$ (frequency for extracting the phase) and the higher oscillatory component around $f_2$ (frequency for extracting the amplitude). If one would filter with a bandwidth smaller than $2f_1$ centered at $f_2$, one would come to the representation [**] and the amplitude of the higher component would be constantly 1, hence independent of the phase of the lower frequency component. However, if one filtered with a bandwidth bigger than $2f_1$, the higher component would be defined as the first summand in the representation [*] and hence show CFC. Therefore, our choice of bandwidth defined how we isolate frequency components, and whether we can observe CFC. This hints at the possibility of obtaining false-positive and false-negative results in cross-frequency analysis depending on the choice of the bandwidth. Importantly, to characterize CFC patterns one usually scans the center frequencies $f_1$ and $f_2$ while keeping the bandwidth of their components fixed. However, a fixed bandwidth is problematic because the critical bandwidth to observe modulation (detecting the sidebands around $f_2$) depends on the value of the lower frequency $f_1$. In particular, this restricts CFC measures to find low modulating frequencies ($f_1$) where sidebands fit within the chosen bandwidth, and thus creates an extreme bias against observing potential phase modulations by higher frequencies (Figure 2). A similar case has recently been raised in [15].

## **Different model/statistical approaches to assess phase-amplitude CFC**

Different methods exist to attack the problem of assessing relationships among rhythmic processes from experimental time series. A possible historical classification of different frameworks is as follows: non-linear systems analysis, Fourier based methods, classical stationary time series analysis, non-parametric approaches, and causal statistical modeling. We include below a brief description of each of the mentioned methods.

Notice that in the main text, we find it more helpful to classify concrete methods for CFC analysis using two almost orthogonal dimensions.

*I) Fourier/wavelet analyses.* These start by projecting the data onto a given basis of oscillatory functions. This first step is indeed common to most of practical approaches to CFC. Fourier/wavelet approaches lack an explicit formulation of a model in terms of equations describing the temporal evolution of relevant variables. However, once a basis or dictionary of functions is selected (oscillatory or not) the projection of data onto such basis amounts to describe the signal as a weighted sum of systems, each of them assuming the dynamics of one basis function. Higher-order spectral quantities, such as bi-coherence, aim to capture correlations between the complex components of an oscillatory basis, and thus, putative CFC relations. The double Fourier transform of a time- or phase-dependent autocorrelation function can also be used to characterize cross-frequency correlations [16]. Unfortunately, higher order functionals are increasingly more difficult to estimate, and multivariate extensions of such approaches are very limited from a practical point of view. The most widely used phase-amplitude CFC measures [2,3] can be thought to measure undirected correlations or dependencies between the frequency components of this class of models.

*II) Classical time series techniques.* They largely refer to regression techniques such as fitting the data into models such as autoregressive moving average processes and its many and important variants both in the time and frequency domain. The strong restrictions of the models (e.g. linearity) are responsible for the extreme data-efficiency and practicality of its multivariate extensions. Models explicitly incorporating cyclo-stationarity (non-stationary components that repeat periodically as used to model seasonal components of financial and geophysical models [17]) can readily serve to quantify certain phase-amplitude CFC effects. Other models exist to regress amplitude variables to non-linear functions of the phase such as the generalized linear models proposed by Penny et al. [18]. Due to the restrictive assumptions of the models, they can easily fall short of the range of interactions actually occurring in complex systems.

*III) Non-linear systems analyses.* These analyses typically assume that a low dimensional dynamical system governs the evolution of variables extracted from measured signals. For example, the Kuramoto model can be used to describe the non-linear interaction between the phases (phase coupling) of coupled oscillators. Coupling terms and other coefficients are then fitted/estimated from the data providing a direct interpretation of the interdependencies between variables. For the Kuramoto example, given the time series of the phases from the measured signals, one can estimate the phase-to-phase coupling coefficients between different oscillatory recordings (see [19] for an efficient implementation of maximum-likelihood inference for Kuramoto networks). Similar models are conceivable for the phase-amplitude problem (Stuart-Landau equations). In general, this class of approaches offers an explicit dynamical model of the relations between different variables extracted from a signal, and can easily account for multivariate descriptions (e.g., by adding more coupling terms), and these approaches are data efficient compared to non-parametric approaches. However, while certain aspects of the model can be justified on theoretical grounds, they usually lack a physiological foundation.

*IV) Non-parametric approaches.* These aim to estimate relationships between variables without assuming any specific structure of the model, which makes them ideal for explorative analysis. Often they make use of probability or state-space based descriptions of variables and their relationships, which are sensitive to all nonlinear order interactions as it is the case with information theory functionals. Transfer entropy for example, aims to capture Wiener/observational causal relations between variables (possibly related to processes at different frequency bands) by quantifying the increase of predictability about the future states of one variable once the information about the present and past states of another variable are included. However, these approaches are typically highly expensive in terms of data, especially in multivariate settings due to the curse of dimensionality. Thus, practical estimators must rely on certain mild assumptions such as the smoothness of probability distributions and the wide-sense stationarity or ergodicity of the time-series [20].

*V) Causal statistical modeling.* It offers a scheme to compare how much evidence the data provide about particular models or hypotheses. The hypotheses are usually formalized by a generative model that can incorporate experimental manipulations plus some measurement functions, which relate relevant physical variables to the magnitudes detected by the experimental apparatus. Such procedures typically rely on Bayesian inference to incorporate prior knowledge and estimate the evidence for the different models once new data has been collected. Dynamical causal models (DCM) are being developed to account for biophysical descriptions of neuronal interactions [21]. Unfortunately a practical approach for inferring phase-amplitude cross-frequency interactions has yet to be developed.

VI) *Other possibilities* include parametric approaches, as for example fitting phase-amplitude histograms to Von Mises distributions [22].

**The transitivity of correlation between phase and amplitude**

In non-linear oscillators, amplitude and phase are intrinsically coupled and thus simultaneously influenced by generic perturbations (e.g. input). Thus, under natural conditions it is almost impossible to selectively modify the amplitude, without changing the phase and vice versa [23]. Moreover, since amplitude and phase can exhibit different susceptibility to perturbations or inertia it is not trivial to infer causal relationships from the timing of their dynamics. In addition, the transitivity of correlations makes it hard to distinguish whether the phase of one process and the amplitude of another are directly or indirectly linked, namely via phase-phase or amplitude-amplitude coupling (Supplementary Figure 8A). For example, the phase of the low frequency component ($\Psi_L$) might influence the phase of the high frequency component ($\Psi_H$) (phase-phase coupling). Since phase and amplitude of the high frequency process are intrinsically coupled, one will observe phase-amplitude coupling. Consequently, it is advisable to partial out indirect ways (e.g. phase-phase coupling) of phase-amplitude modulation in order to assign a functional role to a specific type of coupling. Notice that also some of the vertical couplings in the Supplementary Figure 8A can also appear as a function of the estimators used to define phase and amplitude. For example, the amplitude and phase defined by the analytical signal approach (using Hilbert transforms) are not fully independent and even a nominal change in one of them induces a perturbation in the other (Supplementary Figure 8B).

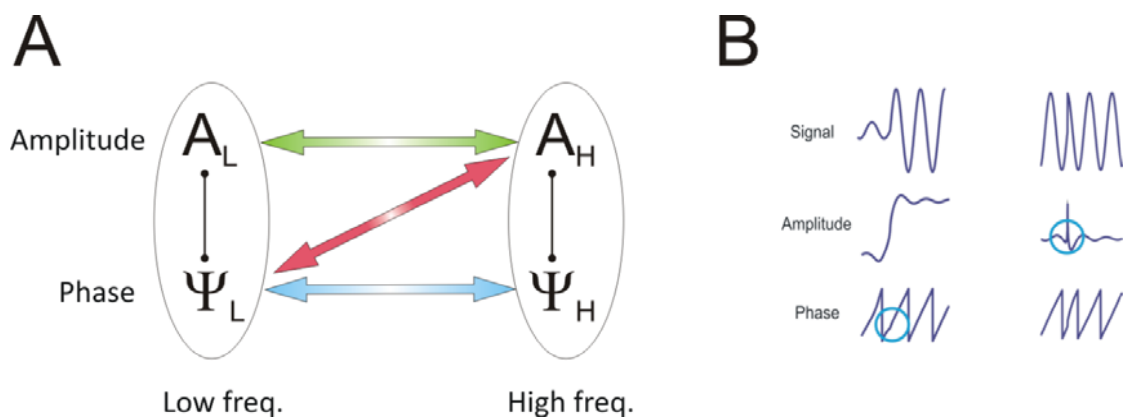

**Supplementary Figure 8. A**. For non-linear oscillators phase and amplitude at the same frequency are intrinsically linked. Therefore, there are different ways to obtain phase-amplitude coupling between a low (L) and a high (H) frequency component of different oscillators. **B.** Estimation of instantaneous phase and amplitude can also couple phase and amplitude dynamics. Nominal changes in either the amplitude (left) or phase (right) of a sinusoidal are simultaneously reflected in both the instantaneous phase and amplitude as obtained by the analytical approach (Hilbert transform).

**Supplementary discussion on causality methods**

As discussed in the main text, approaches for the detection of (observational) causality would improve the analysis of CFC by adding a sense of directionality, thereby constraining the number of possible scenarios for the observation of spurious CFC. Unfortunately, currently applied observational causality approaches meet several difficulties that need to be resolved before they can be applied to interactions between frequency components in neurophysiological data. For example, the frequently used linear Granger formalism is blind to any cross-frequency effects. Some attempts have been made to apply similar techniques to assess linear causality between features of the original signals, i.e. instantaneous phases and amplitudes of different spectral bands. However, these interactions may themselves be mediated by highly non-linear processes and thus remain invisible for the Granger formalism. On the other hand, non-parametric approaches in neuroscience such as transfer entropy [20] account for all orders of non-linear interactions, but typically need long stretches of data for reliable estimation.

In addition, the uncertainty principle of harmonic analysis limits the temporal resolution with which a spectrally resolved component can be localized. By this principle, low-frequency components will be more temporally smeared than high-frequency components, because lower frequency components are usually extracted with a smaller bandwidth. Thus, the onset of low-frequency components will be advanced more than that of high-frequency components if non-causal filtering procedures such as the Hilbert transform are used. Even when causal filters are applied, the problem arises if unequal signal to noise ratios are encountered for different components. This hampers the interpretation of most causality measures, because they rely on temporal order.